\documentclass[fleqn,12pt,twoside]{article}
\usepackage[headings]{espcrc1}

\readRCS
$Id: espcrc1.tex,v 1.2 2004/02/24 11:22:11 spepping Exp $
\usepackage{graphicx}
\usepackage[figuresright]{rotating}

\newcommand{\AmS}{{\protect\the\textfont2
  A\kern-.1667em\lower.5ex\hbox{M}\kern-.125emS}}

\title{Thermalization of gluon matter including $gg \leftrightarrow ggg$
interactions}

\author{A. El\address[MCSD]{Institut f\"ur Theoretische Physik,
        Johann Wolfgang Goethe Universit\"at Frankfurt, \\
        Max-von-Laue Str. 1, D-60438 Frankfurt, Germany}%
        \thanks{Email address: el@th.physik.uni-frankfurt.de},
        C. Greiner\addressmark\thanks{Email address:
        carsten.greiner@th.physik.uni-frankfurt.de}
and Z. Xu\addressmark\thanks{Email address: xu@th.physik.uni-frankfurt.de} }
       
\runtitle{Thermalization of gluon matter}
\runauthor{A. El, C. Greiner and Z. Xu}

\begin{document}

\maketitle

\begin{abstract}
Within a pQCD inspired kinetic parton cascade we simulate 
the space time evolution of gluons which are produced initially in a heavy 
ion collision at RHIC energy. 
The inelastic gluonic interactions $gg \leftrightarrow ggg$ do play an 
important role: 
For various initial conditions it is found that
thermalization and the close to ideal fluid dynamical 
behaviour sets in at very early times. Special emphasis is put on
color glass condensate initial conditions and the `bottom up
thermalization' scenario. Off-equilibrium $3\rightarrow 2$ processes make up
the very beginning of the evolution leading
to an initial decrease in gluon number and a temporary avalanche
of the gluon momentum distribution to higher
transversal momenta.
\end{abstract}

\section{INTRODUCTION}

It had been demonstrated that the measured momentum anisotropy parameter $v_2$
at RHIC energy can be well understood 
if the expanding quark-gluon matter is described by ideal hydrodynamics.
This important finding suggests that 
a strongly interacting and locally thermalized state of matter 
has been created 
which behaves almost like a perfect fluid. 
On the other hand, the initial situation of the 
quark-gluon system is far from thermal equilibrium. It is thus
important to understand how and which microscopic partonic interactions can 
thermalize the system within a short timescale and can be responsible
as well for its (nearly) ideal hydrodynamical behaviour.

A traditional way to study thermalization of particles is to carry out
microscopic transport simulations. 
A standard parton cascade analysis incorporating only
elastic (and forward directed) $2 \leftrightarrow 2$
collisions described via one-gluon exchange,
shows that thermalization and early (quasi-)hydrodynamical behaviour
(for achieving sufficient elliptic flow)
can not be built up or maintained,
but only if a much higher, constant and isotropic cross section 
$\sigma _{eff} \approx 45$ mb
is being employed \cite{MG02}.
The gluons being treated as semi-classical
degrees of freedom would then aquire a collional width 
$\Gamma \approx n \sigma_{eff} v_{rel} $ being
much larger than the typical energy $E \approx 3 T $.
The gluons would then resemble very broad excitations and 
can not be considered at all as
quasi-particles, if such large and constant cross sections
are employed.

In contrast, a kinetic parton cascade algorithm
\cite{xu05} has beeb deneloped
with perturbative QCD inspired processes including
for the first time inelastic (`Bremsstrahlung')
collisions $gg \leftrightarrow ggg $.
The multiparticle back reation channel
is treated fully consistently
by respecting detailed balance within the same 
algorithm.
The three-body gluonic interactions are described by the matrix element
\begin{equation}
\label{m23}
| {\cal M}_{gg \to ggg} |^2 = \left ( \frac{9 g^4}{2} 
\frac{s^2}{({\bf q}_{\perp}^2+m_D^2)^2} \right ) 
\left ( \frac{12 g^2 {\bf q}_{\perp}^2}
{{\bf k}_{\perp}^2 [({\bf k}_{\perp}-{\bf q}_{\perp})^2+m_D^2]} \right )
\Theta(k_{\perp}\Lambda_g-\cosh y) \, ,
\end{equation}
where an effective Landau-Pomeranchuk-Migdal suppression is considered.
$\Lambda_g$ denotes the gluon mean free
path, which is given by the inverse of the total gluon 
collision rate $\Lambda_g=1/R_g$.
This leads to a lower cutoff of $k_{\perp}$ and
to an effective increase of the collision angles.
Typical Debye screened
total pQCD cross section scale roughly like the inverse temperature squared,
$\sim 1/T^2$, so that the characteristic ratio
$\Gamma /E $ should be (significantly) smaller than 1,
as long as the coupling stays small enough.
The possible importance of incorporating inelastic reactions
on overall thermalization was
raised in the so called `bottom up thermalization' picture \cite{baier01}.
It is intuitively clear that gluon multiplication should
not only lead to chemical equilibration, but
also should lead to a faster kinetic equilibration.
This has been demonstrated in detail in \cite{xu05}.

\begin{figure}
\begin{minipage}[t]{75mm}
\includegraphics[scale=0.75]{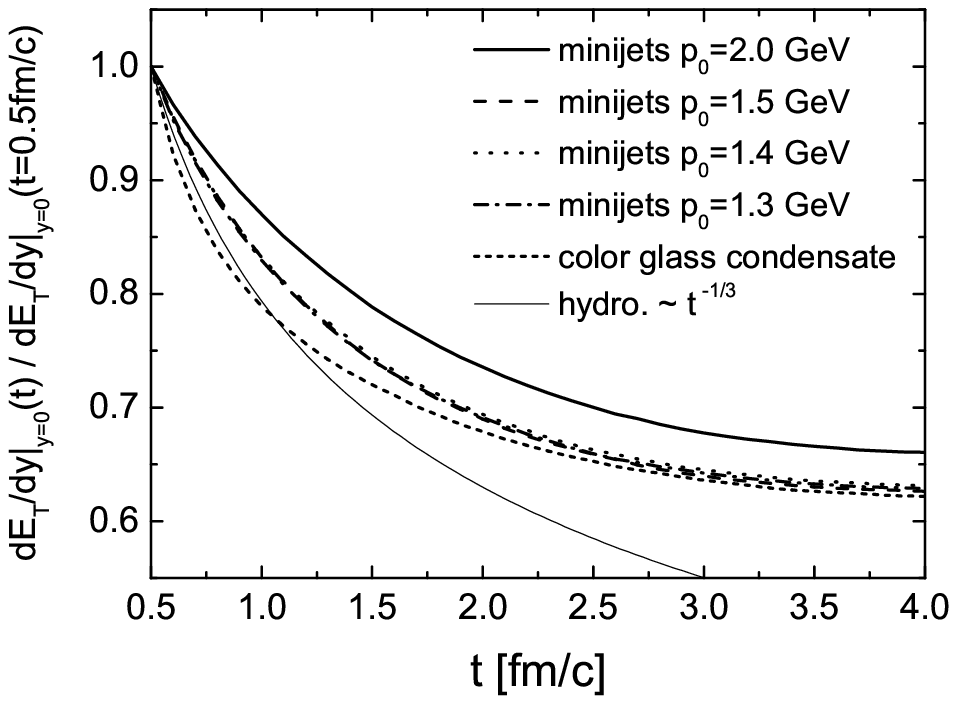}
\caption{ 
Time evolution of nomalized transverse energy 
at midrapidity. Results are obtained from simulations with
different initial conditions of gluons for a central Au Au
collision.}
\label{et}
\end{minipage}
\hspace{1mm}
\begin{minipage}[t]{75mm}
\includegraphics[scale=0.75]{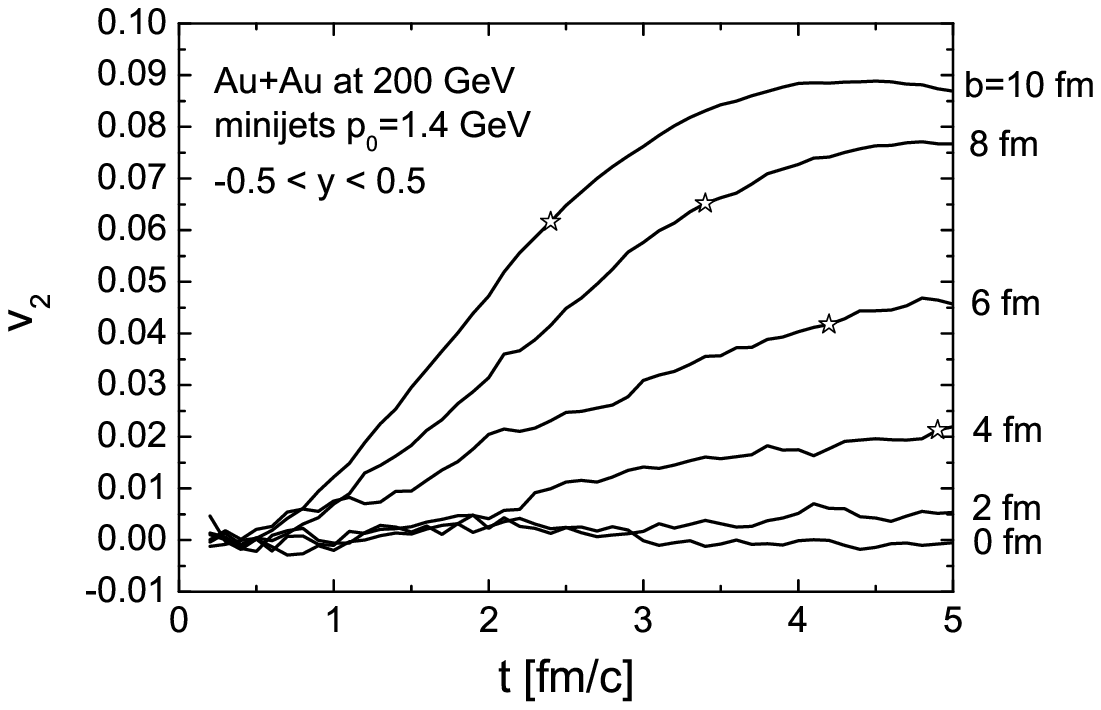}
\caption{ 
Time evolution of the elliptic flow $v_2$ at midrapidity for various
impact parameter $b$.}
\label{v2}
\end{minipage}
\end{figure}

A first picture of the production of the primary partons 
at the very onset of a heavy
ion collision is based on a free superposition of
minijets being liberated in the individual semihard nucleon-nucleon
interactions with transverse
momentum being greater than a certain cutoff $p_0$ \cite{xu05}.
Conditions  of a color glass condensate fot the initial partons 
will also be given and further detailed 
in the next section.

Phenomenologically the cutoff $p_0$ can be chosen in a way to fit
the (final) $dE_T/dy$ as seen in experiment. 
A value of $p_0=1.4$ GeV would meets a final $dE_T/dy \approx 625 $ GeV.
Thus the minijets production at $p_0=1.4$ GeV seems to give an appropriate
initial condition of gluons at RHIC. In Fig. \ref{et} the
transverse energy at midrapidity is depicted,
obtained from simulations with the
different initial conditions, being normalized to
the energies at
$t=0.5$ fm/c.
One sees a rather unique behavior 
of a decreasing, normalized
$dE_T/dy|_{y=0}$, especially for initial conditions employing minijets 
with $p_0=1.3-1.5$ GeV and CGC with $Q_s=1.0$ GeV. 
Comparisons to the normalized transverse energy per
unit rapidity for an ideal hydrodynamical expansion 
show that the collective expansion due to the pQCD 
interactions is quasi ideal at $0.5-1.5$ fm/c. At later times the decrease
of $dE_T/dy|_{y=0}$ slows down, since the collision rates become smaller,
especially in the outer, transversally expanding
region.
\begin{figure}
\begin{minipage}[t]{75mm}
\includegraphics[scale=0.75]{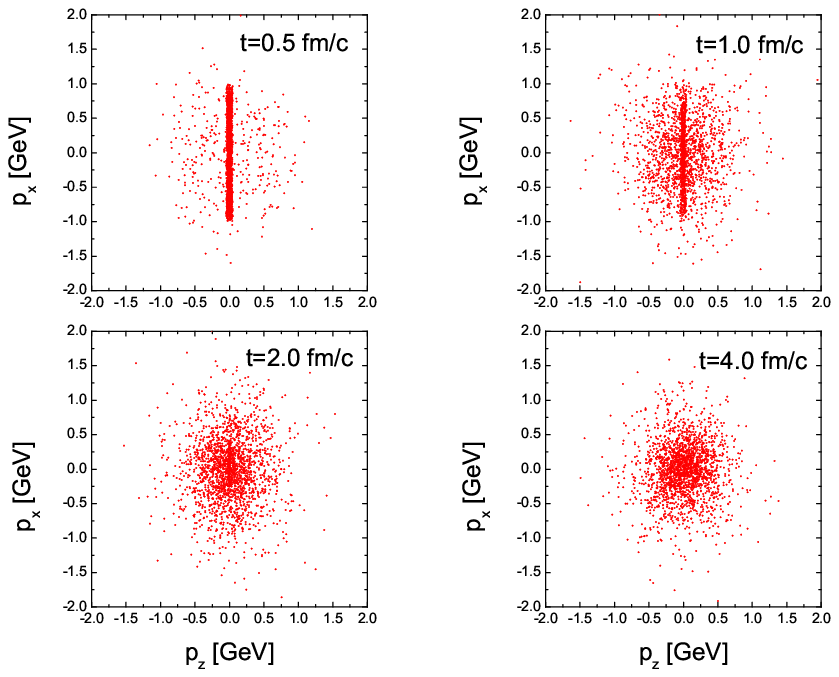}
\caption{ 
The evolution of the local momentum occupation
within  a typical color glass
condensate initial condition taken at $\tau_0=0.4$ fm/c 
is shown for four subsequent times. }
\label{cgc1}
\end{minipage}
\hspace{1mm}
\begin{minipage}[t]{75mm}
\includegraphics[scale=0.75]{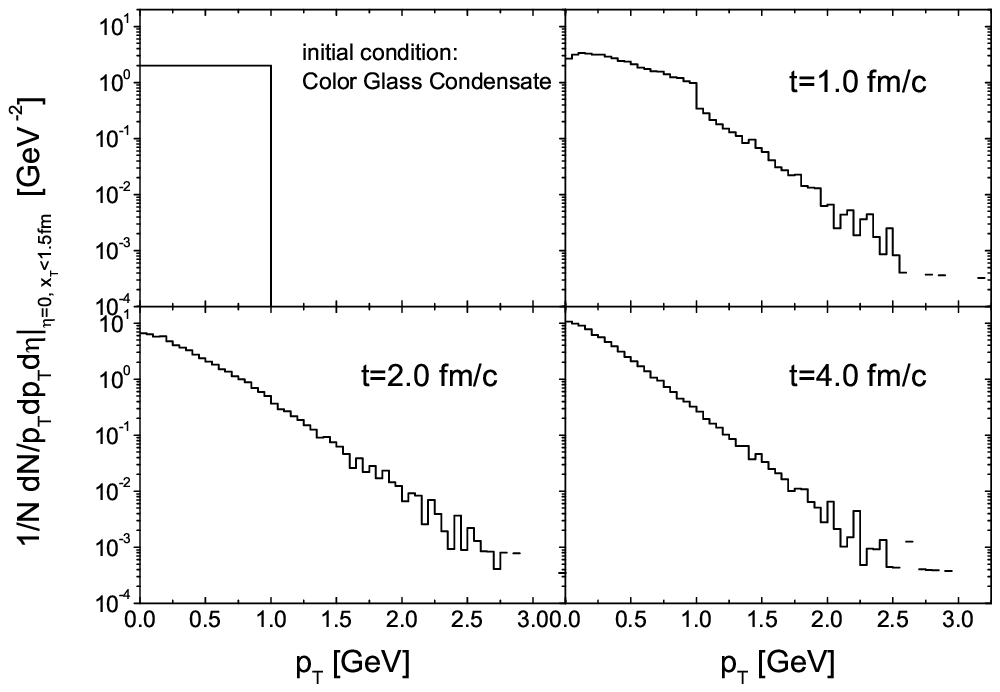}
\caption{ 
Respective transverse momentum spectrum in the central region.}
\label{cgc2}
\end{minipage}
\vspace*{-5mm}
\end{figure}

Taking $p_0=1.4$ GeV for the initial minijets, the parton
evolution for noncentral collisions at RHIC energy is
simulated in order to calculate
the elliptic flow parameter $v_2$. Figure \ref{v2} shows the time
evolution of $v_2$ extracted at midrapidity for various impact
parameter $b$. These calculations are still preliminary.
The results gives 
strong indication that an early pressure is being built up
within that pQCD inspired description. The symbols
in Fig. \ref{v2} mark the time from which the energy density in the
central region decreases below 1 $\mbox{Gev/fm}^3$.
If we take the $v_2$ values at that  marked times as the contribution from 
the partonic phase, they lie well in the region covered
by the experimental data.

\section{Thermalization of the color glass condensate}

For the initial gluon distribution from the saturation picture we employ
an idealized and boost-invariant form \cite{bv01} and
express it as
\begin{equation}
\label{fcgc}
f(x,p)|_{z=0}=\frac{c}{\alpha_s \, N_c}\, \frac{1}{\tau_f}\, \delta(p_z)\,
\Theta(Q_s^2-p_T^2)\,,
\end{equation}
which is described by $Q_s$, the momentum scale at
which gluon distribution saturates. The boost-invariance leads to
the equality of momentum and space-time rapidity, i.e., $\eta=y$,
for the initial gluons. 
The value of parameters are taken 
from \cite{bv01}: $N_c=3$ for SU(3), $c=1.3$, $\alpha_s=0.3$,
and the corresponding formation time, at which the gluon
distribution becomes dilute enough, is given as $\tau_0=c/(\alpha_s N_c
Q_s)$. Hence, the cascade operates from $\tau_0 = 0.4 $fm/c for
$Q_s=1 $ GeV and  $\tau_0 = 0.18 $fm/c for
$Q_s=2 $ GeV and $Q_s=3$ GeV.
For the calculation depicted in Figs. \ref{cgc1} and \ref{cgc2}
the gluons are
produced within a transverse radius of $6$ fm (Au nucleus) and within
 $|\eta| < 3$ longitudinally.
Fig. \ref{cgc1} states a first dynamical realization
of the so called `bottom up thermalization' scenario 
as advocated in \cite{baier01}.  The evolution of the momentum occupation
is shown for four subsequent times sampled 
within a space-time rapidity
interval of $\Delta \eta = 0.1$ and a central transverse region 
of $R\leq 1.5$ fm. One recognizes the population 
of the `soft' gluons and a subsequent degradation of
the `hard' initial gluons. In addition, also harder
gluons are produced. All this happens roughly within the first 1 fm/c.
In the last picture we also see that all particles are clearly
more centered around the origin, demonstrating 
the ongoing cooling and quasi hydrodynamical behaviour from 2 to 4 fm/c.
In Fig. \ref{cgc2}
the transverse momentum spectrum is given for various times. 
Comparing the $p_T$ spectra
between $2.0$ and $4.0$ fm/c,
the local system at the central region clearly is at 
equilibrium at that later times, and
the spectrum steepens continuously, since the thermodynamical 
work in outward direction cools down the system.

\begin{figure}
\begin{minipage}[t]{75mm}
\includegraphics[scale=0.45,angle=270]{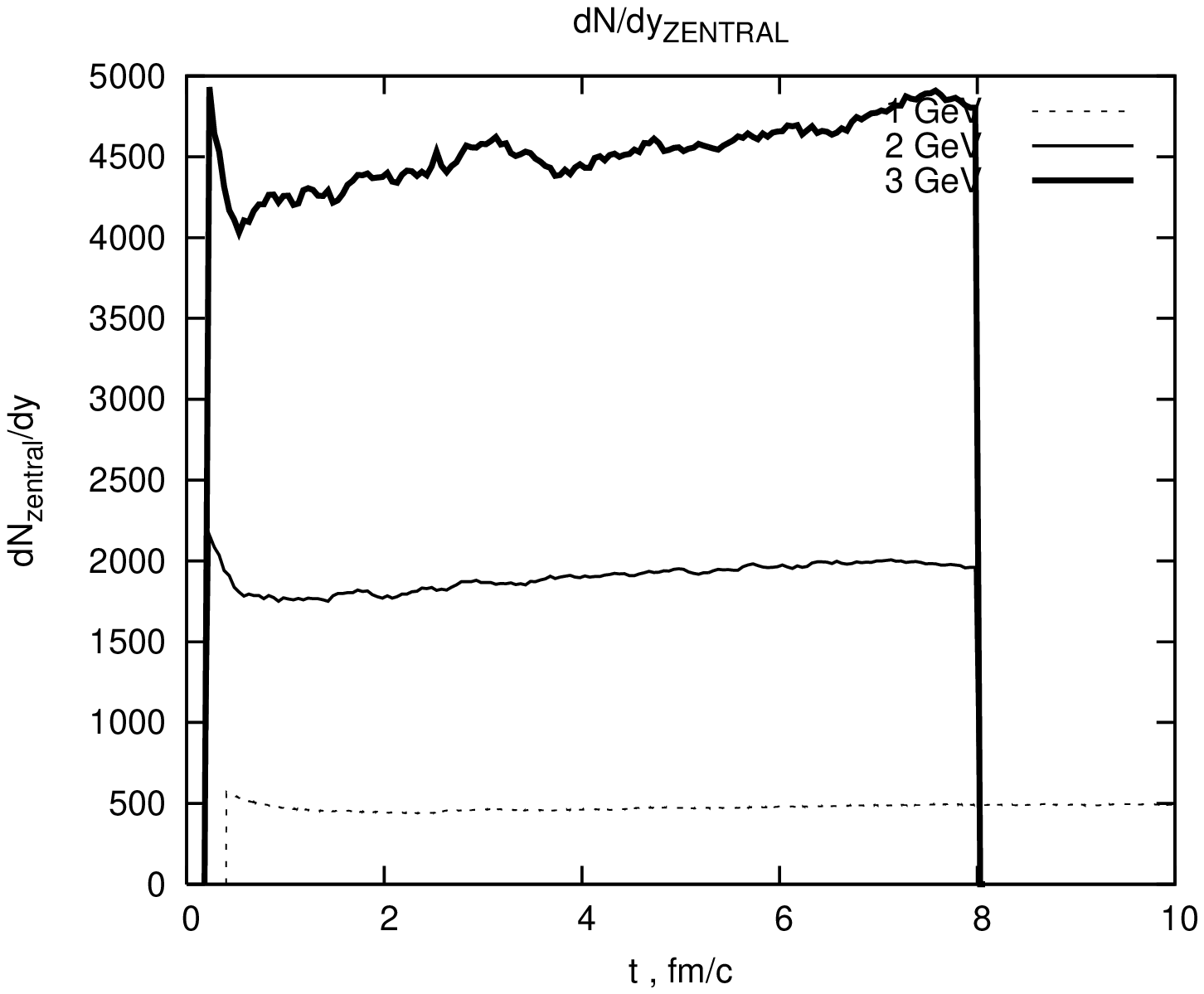}
\caption{
Time evolution of particle number per rapidity
for various saturation momenta $Q_s$.}
\label{dndy}
\end{minipage}
\hspace{1mm}
\begin{minipage}[t]{75mm}
\includegraphics[scale=0.45,angle=270]{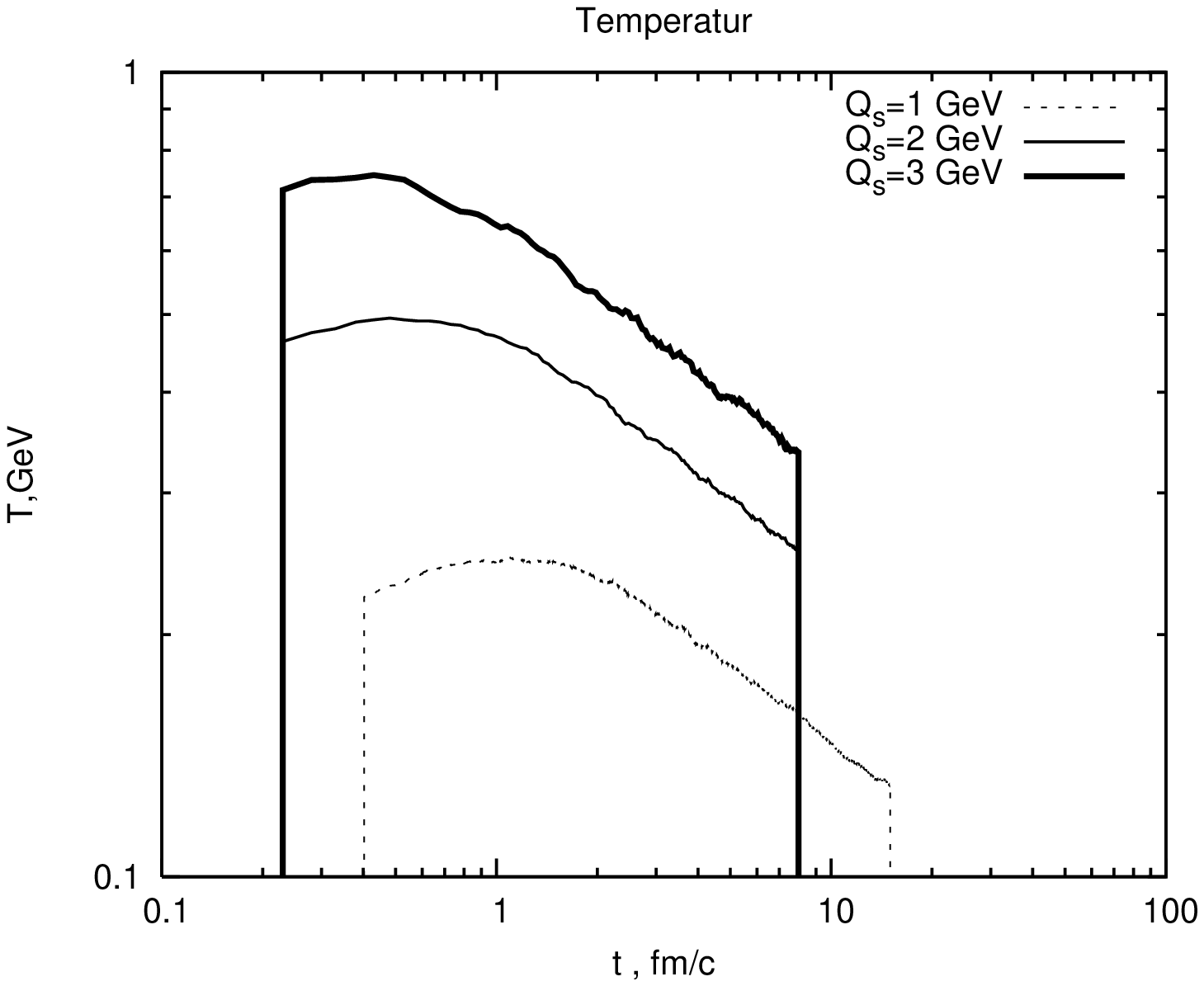}
\caption{
Time evolution of the effective temperature $T=E/3N$.}
\label{temp1}
\end{minipage}
\vspace*{-5mm}
\end{figure}

Thermalization somehow resembles the idealistic `bottom up scenario'
\cite{baier01} with low momentum gluons being produced and populated.
On the other hand, a couple of important differences do show up.
In Fig. \ref{dndy} the time evolution of particle number per rapidity
for various saturation momenta $Q_s$ is shown. In these simulations 
strict Bjorken geometry in longitudinal direction is preserved
by reflection at cylindrical boundaries \cite{el06}.
The number of gluons is initially slightly
decreasing, although a strong parametric enhancement has been advocated
in \cite{baier01} due to Bremsstrahlung production, which is not
observed in the present calculations.
This is true for all chosen saturation scales.
The initial gluons are seemingly 
oversaturated, if the system thermalizes rather immediately \cite{el06}.
Only if thermalization would happen on a much longer time scale,
the gluon number would become undersaturated at a subsequent time,
at which net gluon production would then take place.

Also it is found \cite{el06}
that indeed the $ggg\to gg$ processes 
dominate the dynamics from the very beginning up to
approximately 2 fm/c. This causes, in particular, a production
of gluons with momenta much larger than the saturation scale
$Q_s$. This can also be seen from Figs. \ref{cgc1} and \ref{cgc2}, where
gluons with such high momentum do show up. 
This behaviour resembles an avalanche from low momentum to high momentum, 
which has recently also be found in classical 
Yang-Mills simulation \cite{dn06}. Both phenomena might be connected 
and clearly deserve further investigation.
It also seems that the gluon distribution at higher momenta
above the saturation scale evolves more faster to an exponential
shape than the low momentum gluons \cite{el06}. 
The name bottom-up has then to be reversed.

 \begin{figure}[t]
	\centering
	\includegraphics[scale=0.6,angle=270]{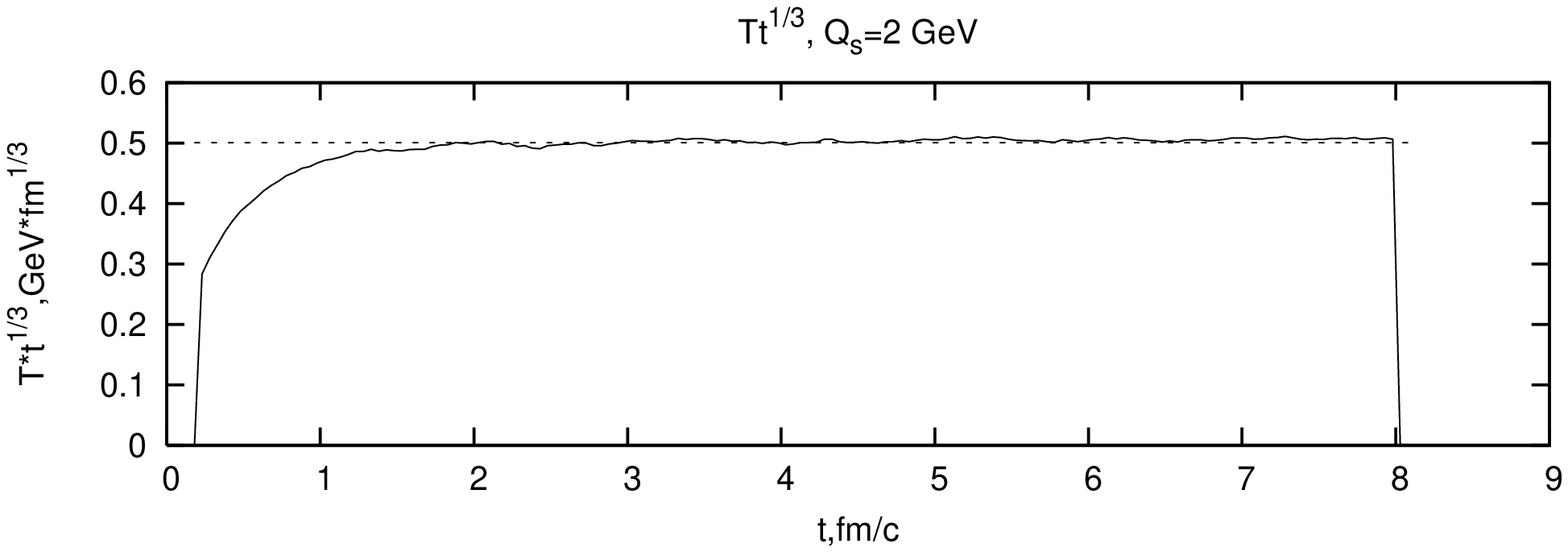}
	\includegraphics[scale=0.6,angle=270]{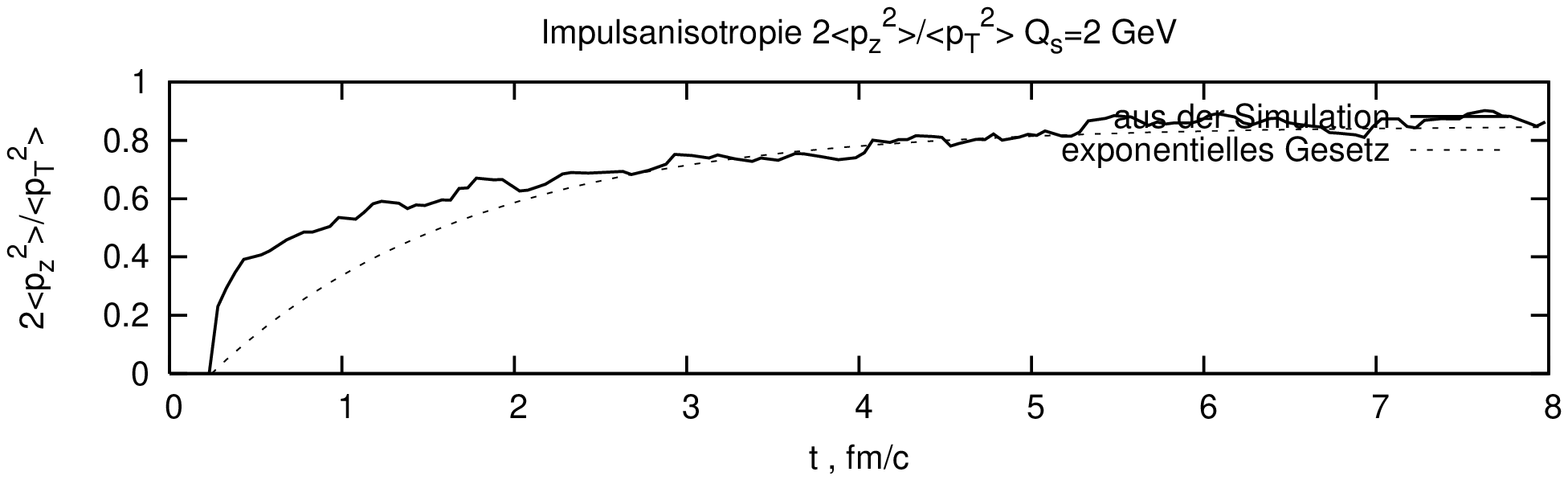}
\caption{
Time evolution of the rescaled temperature 
$T\cdot t^{\frac{1}{3}}$ and the momentum anisotropy.}
\label{Tt}
\vspace*{-5mm}
\end{figure}

As a last point the extraction of a possible thermalization scale
is briefly discussed \cite{el06}. 
Fig. \ref{temp1} gives the time evolution of the effective temperature
$T=E/3N$ for three different saturation scales. At the very beginning 
the temperature increases by a small amount because of the
decrease in gluon number, but then decreases more or less in accordance
with ideal 1-dimensional hydrodynamical expansion.
This can be seen more explicitly in the upper Fig. \ref{Tt}.
(Quasi-)Ideal hydrodynamical behaviour sets in at a timescale
$\theta=3.3~fm/c(Q_S=1 ~GeV)$,  $\theta=1.9~fm/c(Q_S=2 ~GeV)$
and  $\theta=1.2~fm/c (Q_S=3 ~GeV)$. Thermalization should set 
in slightly earlier. 
Kinetic equilibration is characterized by the time evolution of
the momentum anisotropy, $2<p_z^2>/<p_T^2>$ , which maintains a value
of one at equilibrium. The momenta of
gluons are initially directed transversely in local discs according to
Eq. (\ref{fcgc}). Therefore the momentum anisotropy has a vanishing initial
value. An exponential fit to the time evolution gives  
$\theta=2.5~fm/c\, (Q_S=1 ~GeV)$,  $\theta=1.5~fm/c\, (Q_S=2 ~GeV)$
(cf lower Fig. \ref{Tt}),  $\theta=1.0~fm/c \, (Q_S=3 ~GeV)$, 
although equilibration happens faster than exponetial
at the beginning.


\begin{thebibliography}{99}
\bibitem{MG02} D. Molnar and M. Gyulassy, 
Nucl. Phys. A697, 495 (2002).
\bibitem{xu05} Z. Xu and C. Greiner, Phys. Rev. C 71, 064901 (2005);
Nucl. Phys. A 774, 787 (2006); Eur. Phys. J. A 29, 33 (2006).
\bibitem{baier01} R. Baier, A.H. Mueller, D. Schiff, and D.T. Son,
Phys. Lett. B 502, 51 (2001).
\bibitem{el06} A. El, diploma thesis (july 2006), university of Frankfurt,
A. El, Z. Xu and C. Greiner, in preparation.
\bibitem{bv01} J. Bjoraker and R. Venugopalan, Phys. Rev. C 63, 024609 (2001).
\bibitem{dn06} A. Dumitru, Y. Nara and M. Strickland, arXiv:hep-ph/0604149.
\end{thebibliography}
\end{document}